\documentclass[useAMS,usenatbib,letterpaper]{mn2e}

\usepackage{times}

\usepackage{graphicx}
\usepackage{amsmath}
\usepackage{amssymb}
\usepackage{natbib}

\newcommand{\apj}{Ap.J.}
\newcommand{\aj}{A.J.}
\newcommand{\mnras}{MNRAS}

\def\trh0{t_{rh}(0)}

\def\apgt{\ {\raise-.5ex\hbox{$\buildrel>\over\sim$}}\ }
\def\aplt{\ {\raise-.5ex\hbox{$\buildrel<\over\sim$}}\ }

\title[IMBH in old globular clusters]
{Dynamical evidence for intermediate mass black holes in old globular clusters}
\author[Michele Trenti]
{Michele Trenti$^{1}$\thanks{E-mail addresses:
trenti@stsci.edu (MT)}\\
$^{1}$Space Telescope Science Institute, 3700 San Martin Drive, Baltimore, MD, 21218 }

\begin{document}

\date{Submitted to MNRAS Letters 12 Oct. 2006 }

\pagerange{\pageref{firstpage}--\pageref{lastpage}} \pubyear{2006}

\maketitle

\label{firstpage}

\begin{abstract}

We present an indirect dynamical evidence, based on the measure of the
core to half mass radius ratio ($r_c/r_h$), that a significant
fraction of globular clusters in our galaxy with an age greater than
10 half-mass relaxation times ($t_{rh}$) host an intermediate mass
black hole (IMBH). In fact, after a few $t_{rh}$ much of the memory
about the details of initial conditions is erased and $r_c/r_h$ is
determined solely by the balance of the stellar encounters energy
production in the core with the dissipation due to global expansion of
the cluster. Here we compare the observed values of $r_c/r_h$ from a
sample of 57 galactic globular cluster, selected to be dynamically old
and not strongly influenced by tidal forces, with the theoretical
expectation for the $r_c/r_h$ ratio based on analytical models and
detailed numerical simulations. The simulations of the evolution of
star clusters considered include combinations of single stars,
primordial binaries and IMBH. For at least half of the clusters in our
sample the observed $r_c/r_h$ ratio appears to be too large to be
explained without invoking the presence of an IMBH at the center of
the system.

\end{abstract}

\section{Introduction}

The existence and detection for both stellar mass black holes as well
as supermassive black holes are commonly accepted and recognized in
the astronomical community \citep{ho99,kor01}. Intermediate mass black
holes (IMBH), i.e. black holes with mass in the $10^3-10^4 M_{\sun}$
range, possibly formed by runaway mergers of massive stars \citep{pz1}
are expected to fill in the mass range from stellar to supermassive
BH. Indeed, evidence for the presence of IMBH in star cluster is
accumulating, but so far there are only two globular clusters -
\emph{M15} and \emph{G1} - that present observational evidence for the
existence of IMBH \citep{geb02,geb05,vdm02,ger03}. The cases for the
presence of IMBH are based on the challenging determination of the
precise kinematics of the systems, that has to be compared with
detailed dynamical models in order to infer the presence of the
central BH. At the level of velocity dispersion and full line of sight
velocity dispersion profiles, the \emph{direct} dynamical influence of
an IMBH is limited within the sphere of influence of the BH, hard to
resolve observationally (but see the recent high-resolution HST survey
carried out by \citealt{noy06}). In addition, the intrinsic three
dimensional density (and velocity dispersion) cusp around the central
BH is smeared out when the data are projected along the line of sight.

However, even if the direct dynamical signature of a central BH is
hard to detect observationally, the presence of an IMBH at the center
of a collisional stellar systems has profound consequences on the
global dynamics. In particular the visual appearance of a globular
clusters hosting an IMBH is that of a system with a sizeable core
\citep{bau05,tre06b} and not that of very concentrated cluster, as
could be naively inferred from the existence of the Bahcall-Wolf
$1/r^{1.7}$ density cusp \citep{bah76} within the sphere of influence
of the BH. This is because the presence of a central BH and of its
compact density cusps generates, through stellar encounters, a
sufficient amount of energy to halt the core collapse, if the initial
conditions start with a wide core, or to fuel a core expansion if the
initial configuration is centrally concentrated. In \citet{tre06b} we
have shown that star clusters with a central IMBH starting from a
variety of initial stellar density profiles (i.e. King models with
concentration parameter $W_0=3,5,7,11$ as well as Plummer models) all
evolve toward a common value of the core to half mass radius ratio
($r_c/r_h \approx 0.3$) on a timescale of a few initial half-mass
relaxation times. This is indeed not surprisingly: a similar behavior
is observed when primordial binaries are the responsible driver for
energy production in the center of the cluster \citep{tre06a}, in
agreement with the theoretical expectation formulated by
\citet{ves94}. As binaries alone are less efficient at producing
energy than an IMBH, in this case we have $r_c/r_h \lesssim 0.06$
(where the upper limit is reached in presence of a number density of
at least $50\%$ binaries in the core) for a typical globular
cluster. Only when no or little binaries are present there can be a
deep core collapse, ending when $r_c/r_h \lesssim 0.02$ (see,
e.g. \citealt{heg06}).

Given this theoretical framework, the observed core to half mass
radius ratio appears indeed a powerful indicator of the dynamical
configuration of the core of an ``old'' globular cluster, i.e. a
globular cluster that has evolved for about $5$ to $10$ half-mass
relaxation times. $r_c/r_h$ is in addition a relatively easy
photometric measurement, that can be applied even to extra-galactic
globular clusters, without worrying about the need of a very accurate
measure. In fact, the combination of numerical simulations and
theoretical modeling predicts that this ratio increases by about a
factor $3$ moving from clusters with single stars only to
clusters with a sizeable population of binaries; when an IMBH is
present $r_c/r_h$ is $4$ to $5$ times larger than in the case when
only binaries are present and $12$ to $15$ times larger than when the
cluster contains single stars only.
 
In this letter we present a preliminary application of the use of
$r_c/r_h$ as an indirect indicator for the presence of an IMBH in the
cores of 57 galactic globular clusters that satisfy a conservative
dynamical age criterion, i.e. their half-mass relaxation time is
shorter than $10^9 yr$. In particular, we show that the observed core
to half mass radius ratio in more than half of the clusters in the
sample is so large that the only general dynamical explanation can be
given in terms of the presence of a central BH. The paper is organized
as follows: in Sec.~\ref{sec:picture} we summarize the current
numerical and theoretical understanding of the evolution of $r_c/r_h$,
in Sec.~\ref{sec:data} we present the sample of galactic globular
clusters that we use in this analysis and we discuss their properties
in terms of our theoretical expectations. Finally we conclude in
Sec.~\ref{sec:conc}.

\section{Global evolution of a cluster with binaries and IMBH}\label{sec:picture}

\begin{figure}
\resizebox{\hsize}{!}{\includegraphics{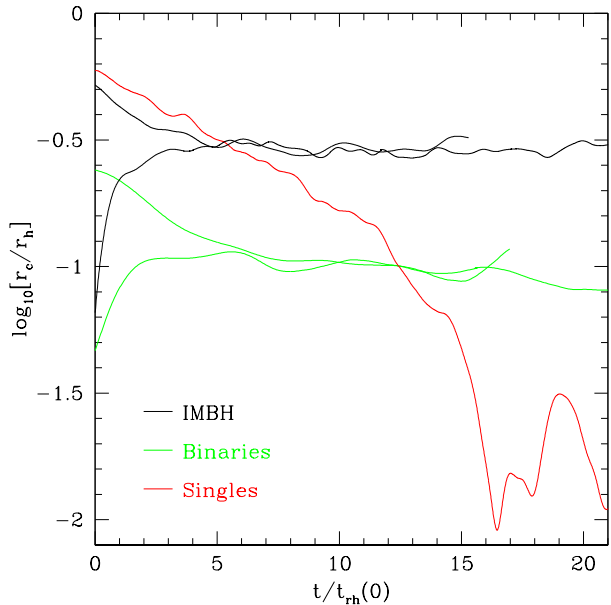}}
\caption{Evolution of $r_c/r_h$ in a series of runs with equal mass
stars from \citet{heg06,tre06a,tre06b} starting from a variety of
initial conditions that include single stars only (red line: Plummer
model, N=8192), primordial binaries with galactic tidal field (green
line: King model $W_0=7$ \& $10\%$ binaries, $W_0=11$ \& $20\%$
binaries, $N=16384$) and with IMBH (black line: $m_{bh} = 1.4\%
M_{tot}$ \& $10\%$ binaries, Plummer model, N=8192; $m_{bh} = 3\%
M_{tot}$, no binaries, King model $W_0=11$, N=4096). When a realistic
mass spectrum is considered for single stars runs the core collapse is
much faster ($\approx 2~t_{rh}$, see
\citealt{che90}).}\label{fig:rcrh_th}
\end{figure}

As discussed in quantitative detail in a series of paper devoted to
the dynamical evolution of star clusters with primordial binaries and
IMBHs \citep{heg06,tre06a,tre06b}, the long term evolution of a
globular cluster is driven by the heat flow generated in the core of
the system toward the halo (see also \citealt{ves94}). Three main
processes can be identified at the base of this heat flow, that
regulates the evolution of $r_c/r_h$ (shown in Fig.~\ref{fig:rcrh_th}
for representative simulations of our numerical simulations program).
(1) When only single stars are present the core behaves as a
self-gravitating systems with negative specific heat (e.g. see
\citealt{heg03}) and a thermal collapse happens leading to a core
contraction. The core contraction proceeds for about $10-15$ half mass
relaxation times for equal mass particles, while it is significantly
faster when a mass spectrum is considered \citep{che90}. The collapse
lasts until the central density is so high that a few binaries are
dynamically formed by three body encounters. Given the high central
density these binaries interact efficiently with single stars and
become progressively tighter and tighter providing enough energy to
halt the collapse and create a core bounce. When the number of stars
in the system is large enough ($N \gtrsim 10^4$) binary activity may
cause a temporary temperature inversion in the core and subsequent
core expansion. The process is eventually repeated in a series of
``gravothermal oscillations'' \citep{sug83} with the core radius
oscillating around a value roughly $2\%$ of the half mass radius. (2)
When a sufficient fraction (i.e. greater than a few percent) of stars
in the cluster has initially a companion with typical separations
below $10 AU$ (that is ``hard binaries'' that have a binding energy of
at least a few times the total energy of the cluster), energy
generation due to existing binaries is much more efficient than that
due to dynamically formed pairs. Here the equilibrium size of the core
is larger than when only single stars are present. The system may even
show a core radius expansion if the initial conditions are too
concentrated. The precise value of the equilibrium core radius has a
moderate dependence on the number of stars in the systems and on the
number density of binaries in the core: $r_c \approx 5\%~ r_h$ for a
typical globular cluster ($N=3 \cdot 10^5$) with about $40\%$ of
binaries in the core \citep{heg06,tre06a,fre06,ves94}. This
quasi-equilibrium evolutionary phase lasts until there is a sufficient
number of binaries in the core, that is until the binary fraction in
the cluster is reduced at a few percent level. In fact, binaries in
the outer part of the cluster tend to sink toward the core for mass
segregation, so the core binary fraction is replenished until the
total binary fraction is almost exhausted. (3) When an IMBH with mass
of the order of $1\%$ of the total mass of the cluster is present, a
yet more efficient energy production channel is available in the
center of the system via dynamical interactions within the influence
sphere of the BH \citep[][see also
\citealt{bau04a,bau04b}]{tre06b}. The equilibrium value of the core
radius in this case is significantly larger: $r_c/r_h \approx 0.3$
independently from the initial binary fraction. Even in this case
there is an expansion of the core radius on a relaxation timescale if
the initial conditions are too concentrated.

This general picture has been obtained with somewhat idealized N-body
simulations that employed single particles only and did not take into
account stellar evolution. Therefore it is important to assess if the
results that we obtain can be biased by our assumptions. Modest
deviations from our numerical expectations will not impair the
usefulness of $r_c/r_h$ as diagnostic tool to infer the presence of
IMBH, as $r_c/r_h$ is expected in this case to be larger by a factor
$4$ to $5$ with respect to the case where only binaries are present.

To ensure that no major bias is present we examine critically the
idealizations in the numerical runs:
\begin{enumerate}
\item {\emph{Mass spectrum.}} The choice to use equal mass stars is
probably the most significant simplification introduced in our
simulations. When a realistic mass spectrum is present, massive stars
segregate on a relaxation timescale toward the core of the system and
speed up the core collapse by about one order of magnitude, i.e. the
collapse takes about $1-2~t_{rh}$ \citep{che90}. In principle, the
equilibrium value of the core radius could be changed
significantly. However simulations of the dynamics of star clusters
with primordial binaries that include a realistic mass spectrum have
been recently performed with a Monte Carlo code by \citet{fre06} and
their results on $r_c/r_h$ are fully consistent with our more
idealized runs. Similarly, realistic simulations of globular clusters
with an IMBH that include a mass spectrum and stellar evolution (but
without binaries) have been carried out by \citet{bau04b} and again
the presence of a large core radius is confirmed. The analysis of the
final configuration at $t=12 Gyr$ of two largest runs by
\citet{bau04b} - with $N=131072$ - leads to $r_c/r_h \approx 0.27$ for
the run with a \citet{kro01} IMF truncated at $100~M_{\sun}$ and
$r_c/r_h \approx 0.35$ when the IMF is truncated at $30~M_{\sun}$.

\item{\emph{Stellar evolution.}} Our runs do not take into account the
effects of stellar evolution, that limits the lifetime of massive
stars and may lead to the coalescence of tight binaries. These effects
may possibly lead to a more rapid depletion of the binary population
(see \citealt{iva05}), so that a star cluster would burn out its
binary reservoir earlier than expected from our simulations. If this
is the case, then the core undergoes a deep core contraction, like in
the case where only single stars are present. The direction of the
evolution is fortunately in the right direction to avoid an
observational bias when we are interested, as in this paper, to focus
on \emph{large} values of the core to half mass radius. For
simulations with an IMBH, the study of \citet{bau04b} (see point
above), that included stellar evolution, guarantees instead that the
bias due to stellar evolution is not important (see also Fig.1 in
\citealt{bau04b}, where the Lagrangian radii, that is the radii
enclosing a fixed fraction of the total mass of the system, expand
steadily starting from a $W_0=7$ model).

\item{\emph{Number of particles.}} Our direct simulations have been
limited by CPU speed to employ only up to $\approx 20000$
particles. The typical number of stars in globular clusters is larger
by at least one order of magnitude. Given our choice to use equal mass
particles and no stellar evolution, our results are scale-free and can
be adapted to any physical value for the mass and radius scales. From
theoretical considerations by \citet{ves94} when primordial binaries
are present $r_c/r_h$ is expected to have a modest dependence on the
number of particles $N$, that is $r_c/r_h \propto
1/log(0.11N)$. Simulations with $N$ from $512$ to $16384$ indeed
verify that the scaling is approximately consistent with the
\citealt{ves94} model \citep{heg06}. If anything $r_c/r_h (N)$ seems
to decrease slightly \emph{faster} than expected. In addition the
simulations by \citet{fre06} employ a realistic number of particles
and their core radii are consistent with our results extrapolated
using the \citet{ves94} model. When a IMBH is present we do not expect
a significant scaling of $r_c/r_h$ with $N$ as confirmed by our 
analysis of the two $N=131072$ runs of \citet{bau04b}. Even if a
$\log{N}$ scaling were to be assumed despite the indication from the
simulations by \citet{bau04b}, the core radius value extrapolated from
our runs with a IMBH would still be such that $r_c/r_h \gtrsim 0.15$
for a realistic number of particles.

\item {\emph{Tidal field.}} The runs with an IMBH in \citet{tre06b} do
not take into account the presence of a tidal field. This is however
not expected to introduce a significant bias: $r_c/r_h$ in runs with
primordial binaries varies at the $10\%$ level with or without
inclusion of the galactic tidal field \citep{tre06a}.

\item{\emph{Spherical symmetry.}} Our initial conditions are all
spherically symmetric. We do not expect this to be a problem as
globular clusters are typically very close to spherical symmetry. In
addition, the precise details of the starting configuration are not
important on a collisional timescale. To be sure, in the analysis of
the next section we analyze a subsample of the clusters selected by
excluding objects that departs from spherical symmetry, obtaining the
same $r_c/r_h$ distribution as in the parent sample.

\item {\emph{Rotation.}} The simulations that we consider have zero
net angular momentum, i.e. no rotating model is studied. Observational
evidence of rotation in globulars is weak and the velocity dispersion
tensor displays a mild anisotropy at most \citep{mey97}. Therefore it
would be very surprinsing if the comparison with observations turns
out to be strongly biased by not including rotation in simulations.

\end{enumerate}

\subsection{Core radius definition}

Finally, before moving on to compare our theoretical expectations for
$r_c/r_h$ with the data, we have to consider a yet different source of
uncertainty. Numerical simulations give us mass-defined core and
half-mass radii, while luminosity-based quantities are derived from
observations. A discrepancy can therefore arise (1) if the luminosity
profile does not track accurately the mass profile and/or (2) if the
core radius is differently defined in the two cases. 

For the first issue we have to consider that the luminosity of a
globular cluster is dominated by stars in the giant branch, that are
more massive than the average star of an old globular cluster. The
most luminous stars are more centrally concentrated due to mass
segregation than the average, so if a bias is introduced, it goes in
the direction of reducing the observed core radius with respect to a
mass weighted definition. However \citet{che90} derive only modest
color gradients due to this effect. In any case it does not affect the
interpretation of large observed cores. 

The second issue, that is a difference between the observational and
theoretical core radius definition is also not expected to be a major
problem. In our simulations we adopted the density weighted definition
of the core radius from \citet{cas85}, Eq. IV.3:
$$
r_c \equiv \frac{\langle |\vec{x}| \rho \rangle_M}{\langle \rho \rangle_M} = \frac{\sum_i r_i \rho_i m_i}{\sum_i \rho_i m_i},
$$ 
where the sum is carried over all the stars in the simulation, $r_i$
is the distance of the i-th star from the center of the system, $m_i$
its mass and $\rho_i$ the stellar density computed at the star
position. This definition for $r_c$ is closely aligned to the standard
observational practice of defining the core radius as the radius
$r_{\mu}$ where the luminosity surface density has dropped to half its
central value \citep{cas85}.

\section{Data sample}\label{sec:data}

As a preliminary application of the $r_c/r_h$ diagnostic to infer the
presence of an IMBH, we consider the compilation of galactic globular
clusters properties by \citet{har96}, revised in February 2003. For
each globular cluster in the tables of \citet{har96} we extract the
following quantities: (1) core radius ($r_c$); (2) half-mass radius
($r_h$); (3) tidal radius ($r_t$); (4) half-mass relaxation time
($t_{rh}$); (5) ellipticity ($e$); (6) cluster luminosity - i.e. the
absolute visual magnitude - ($M_{Vt}$).

We then proceed to build a uniform and homogeneous data set that
includes only \emph{old} globular clusters to ensure that these
systems are well relaxed by two body encounters. As a first step we
therefore exclude from the analysis:
\begin{enumerate}
\item All the globular clusters for which one of these values is not
quoted (except for $e$);
\item All the globular clusters with $t_{rh}>10^9 yr$. Assuming a
typical globular cluster age of $10^{10}yr$, this ensures that all the
objects in the sample are at least 10 half mass relaxation times
old. Most of the clusters excluded from the sample fail this test;
\item All the cluster that are under the influence of a strong tidal
field effects, i.e. those for which $r_t/r_h<4$. This last selection
criterion is mainly a precaution to exclude systems where $r_c/r_h $
may start fluctuating as $r_t$ approaches $r_h$ (see
\citealt{tre06a}). Only four clusters that pass the previous
selections are excluded due to a small tidal radius.
\end{enumerate}

\begin{figure}
\resizebox{\hsize}{!}{\includegraphics{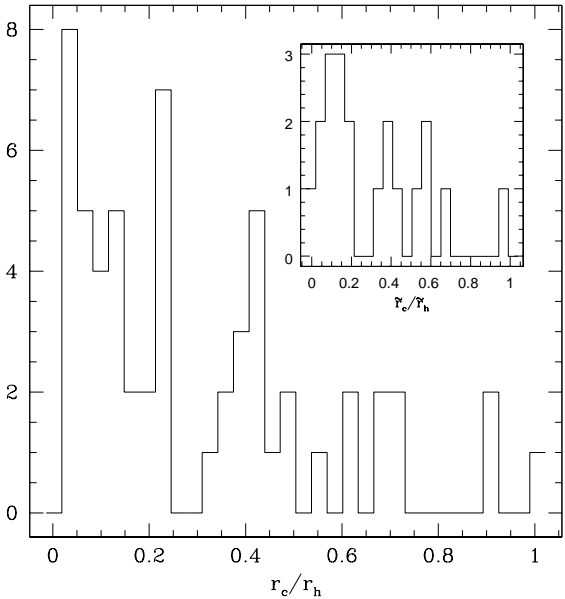}}
\caption{Main panel: distribution of the $r_c/r_h$ ratio in the main
sample of 57 \emph{old} galactic globular clusters (selected to have
$t_{rh}<10^9 yr$). Small panel: $\xi \cdot r_c/r_h$ (corrected to
rescale it to a reference number of particles $N_{ref}=3 \cdot 10^5$ )
for a sub-sample (20 objects) of the main sample, selected by imposing
$t_{rh}<5 \cdot 10^8 yr$ and $e<0.1$. Both panels show a
significant number of clusters with $r_c/r_h \gtrsim 0.2$.
}\label{fig:rcrh_obs}
\end{figure}

After applying our selection criteria we are left with a sample of
$57$ old globular clusters, whose $r_c/r_h$ distribution is plotted in
Fig.~\ref{fig:rcrh_obs}. The plot is extremely interesting: it is
immediately apparent that only a modest fraction of the globular
clusters is characterized by $r_c/r_h<0.1$, that would be consistent
with the core size expected when the system is populated only by
single and binary stars. The majority of the objects in the sample has
a large core ($r_c/r_h > 0.2$) and $r_c/r_h \gtrsim 0.5$ is not
infrequent. The average value of the ratio is $\langle r_c/r_h \rangle
= 0.31$. The median of the distribution is at $r_c/r_h=0.28$. From the
dynamical considerations of Sec.~\ref{sec:picture}, only the presence
of an IMBH appears to be consistent with such large core radius
values.

In Sec.~\ref{sec:picture} we have seen that, when the core to half
mass radius ratio is set by the burning of binaries there is a modest,
logarithmic N-dependence on this ratio, that is reasonably well
captured by the \citet{ves94} model, especially in presence of a
galactic tidal field (see \citealt{tre06a}). To take into account this
factor in the analysis we assume (1) that the total luminosity is a
proxy for the total mass of the system and (2) that the average
stellar mass in each cluster is constant and equal to
$0.7M_{\sun}$. The mass-luminosity relaxation has been checked and
calibrated using the data from \citet{gne97}. We used
$\log_{10}(M_{tot}/M_{\sun}) = A + b \cdot M_{Vt} $ with $A=2.5$ and
$b=-0.38236$. We have then corrected the ratio $r_c/r_h$ by applying a
correction factor $\xi$
$$ 
\xi = \frac{\log{(0.11/0.7~M_{tot}/M_{\sun})}}{\log{(0.11 N_{ref})}}
$$ 
so as to obtain a ratio $\tilde{r_c}/\tilde{r_h} = \xi r_c/r_h$
equivalent to that of a cluster with $N_{ref}=3 \cdot 10^5$, a
standard value in numerical simulations of globular clusters. The
results shown in Fig.~\ref{fig:rcrh_obs} do not change significantly
when this correction is applied: the average value is $\langle
\tilde{r_c}/\tilde{r_h} \rangle = 0.27$ and the median value of the
distribution is $0.22$, only marginally smaller than without the
correction. The presence of an IMBH is therefore still needed to
explain the large value of the core radius for at least half the
objects in the sample.

Could this conclusion be biased by the fact that the observed globular
clusters are for some (improbable) reasons not yet well relaxed, so
that the equilibrium value for $r_c/r_h$ has not yet been reached
after $\approx 10~t_{rh}$? To address this point we enhance our
selection algorithm to select only objects that are at least
$20~t_{rh}$ old, i.e. with $t_{rh}<5 \cdot 10^8 yr$. This reduces our
sample to 25 clusters whose average core radius is still $\langle
r_c/r_h \rangle = 0.36$ (marginally larger than the whole sample
value), while the median is at $r_c/r_h = 0.21$ (marginally smaller
than the whole sample analysis). If the \citet{ves94} $log(0.11N)$
correction is applied we have $\langle \tilde{r_c}/\tilde{r_h} \rangle
= 0.28$ and median value of $\tilde{r_c}/ \tilde{r_h} = 0.17$. The net
effect of the correction is to reduce the mean and the median values,
as by requiring $t_{rh} \leq 5 \cdot 10^8 yr$ we are biased toward
selecting smaller clusters. Even for this reduced sample more than
half of the objects have $\tilde{r_c}/\tilde{r_h} > 0.15$, so the need
for a central IMBH remains strong.

As an additional test to ensure absence of observational biases, we
have further refined the reduced sample with $t_{rh}<5 \cdot 10^8 yr$
to exclude globular clusters that depart from spherical symmetry,
i.e. those clusters with a quoted $e > 0.1$, where the ellipticity
ratio $e$ is defined in terms of the axis ratio ($e=1-a/b$). This
leaves 20 objects in the sample and does not change the results:
$\langle r_c/r_h \rangle = 0.38$ (median $0.23$), while $\langle
\tilde{r_c}/\tilde{r_h} \rangle = 0.29$ (median $0.20$).

All the progressively selective cuts that we applied to the galactic
globular cluster system provide a consistent picture where about half
of the objects have a significantly large core radius, i.e. $r_c/r_h
\gtrsim 0.2$. If an IMBH turns out to be absent in these systems, then
a different dynamical explanation for their large core radii must be
found. If this is indeed the case for the whole sample then a major
rethinking of our knowledge of the dynamics of globular clusters is
needed. Nevertheless a few individual objects in the sample may turn
out to have a large core radius for peculiar reasons, especially if
they have been strongly perturbed and are therefore not as dynamically
old as their relaxation time would imply. One example of a peculiar
object that passes all our selection criteria is the cluster with the
largest $r_c/r_h$ ratio, Pal 13, which is considered a very unusual
object, either due a strong tidal heating during the last
perigalacticon passage or due to the presence of a sizeable dark
matter halo \citep{cot02}.

\section{Conclusions}\label{sec:conc}

In this paper we propose the use of the observed ratio of the core to
half mass radius as a powerful indirect dynamical indicator for the
presence of an IMBH at the center of old stellar cluster. A number of
theoretical and numerical investigations combined together
\citep{ves94,bau04a,bau04b,fre06,heg06,tre06a,tre06b} strongly support
the idea that after about $5$ to $10$ half-mass relaxation times the
value of $r_c/r_h$ in a globular cluster assumes significantly
different values depending only on whether the core contains single
stars ($r_c/r_h \approx 0.02$), a large fraction of hard binaries
($r_c/r_h \approx 0.05$) or an IMBH ($r_c/r_h \approx 0.3$). The
details of the initial conditions, such as the initial density
profile, do not affect this quantity. By analyzing the distribution of
$r_c/r_h$ in a sample of $57$ galactic globular clusters, selected to
ensure that they are well collisionally relaxed, we conclude that
there is a strong indirect evidence for the presence of an IMBH in at
least half of the objects in the sample.

A globular cluster that hosts an IMBH represents a laboratory where
dynamical interactions between hierarchical systems take place
frequently, leading in particular to the formation of triples (and
even quadruples) systems and to the ejections of a number of high
velocity stars, with some able to reach ejection velocities up to
several hundreds km/s (see \citealt{tre06b}).

Intriguingly, some objects in the sample have a core radius that could
possibly be too large ($r_c/r_h \gtrsim 0.5$) to be explained by the
presence of a \emph{single} IMBH. It could therefore be possible that
some of these globular clusters host a binary IMBH. In this case,
however, the heating may be so efficient (e.g. see \citealt{yu03} for
an estimate of the interaction rate of a binary BH with single stars)
that it is not clear whether the system is able to survive for even a
few relaxation times. Therefore an accurate numerical modeling is
critically required before any conclusion about this speculation can
be drawn.


Our conclusions on the presence of singles IMBHs appear instead to be
robust. We have discussed a number of possible biases in the
comparison of numerical simulations with the observations, but while
some of the idealizations introduced in the modeling may induce
limited changes in $r_c/r_h$, we could not identify a possible major
bias. An error of the order at least $300\%$ would be required to be
able to explain the observed distribution of $r_c/r_h$ without
invoking the systematic presence of an IMBH in half of the objects in
the sample.

Clearly more detailed numerical simulations will be required to
evaluate and confirm the presence of IMBH in specific clusters of the
sample. In particular it would be extremely interesting the conversion
of N-body snapshots into synthetic observations that could therefore
be directly analyzed like images acquired by a telescope, avoiding an
indirect comparison between theoretical and observed quantities.

\section{Acknowledgments}
I am very grateful to Holger Baumbardt for providing the snapshots of
his numerical simulations whose analysis is used in
Sec.~\ref{sec:picture}. It is a pleasure to thank Douglas Heggie, Piet
Hut and Massimo Stiavelli for their helpful comments and suggestions.
This work was partially supported by NASA through grant HST-AR-10982.



\label{lastpage}

\end{document}